\documentclass[conference,a4paper,twoside]{IEEEtran}

\usepackage{cite}
\usepackage{amsmath,amssymb,amsfonts}
\usepackage{graphicx}
\usepackage{textcomp}
\usepackage{xcolor}

\usepackage[hidelinks]{hyperref}
\usepackage{xurl}

\usepackage{xspace}
\usepackage{algorithm2e}
\usepackage{cleveref}
\usepackage{array}
\usepackage{enumitem}

\newcommand\concat{%
    \,||\,%
}%

\newcolumntype{L}[1]{>{\raggedright\let\newline\\\arraybackslash\hspace{0pt}}m{#1}}
\newcolumntype{E}{>{\(}r<{\)}}
\newcolumntype{V}{>{\ttfamily}c}

\makeatletter
\let\old@ps@IEEEtitlepagestyle\ps@IEEEtitlepagestyle
\def\confheader#1{%
    \def\ps@IEEEtitlepagestyle{%
        \old@ps@IEEEtitlepagestyle%
        \def\@oddhead{\strut\hfill#1\hfill\strut}%
        \def\@evenhead{\strut\hfill#1\hfill\strut}%
    }%
    \ps@headings%
}

\confheader{%
        \parbox{20cm}{Accepted for publication in: 2024 IEEE Region 10 Conference (TENCON), Singapore, 2024}
}

\makeatother

\begin{document}

\title{Building Touch-Less Trust in IoT Devices}

\author{\IEEEauthorblockN{Steve Kerrison}
\IEEEauthorblockA{\textit{School of Science and Technology} \\
\textit{James Cook University}\\
Singapore \\
\href{mailto:steve.kerrison@jcu.edu.au}{steve.kerrison@jcu.edu.au}}
}

\maketitle

\begin{abstract}
Trust mechanisms for Internet of Things (IoT) devices are commonly used by manufacturers and other ecosystem participants. However, end users face a challenge in establishing trust in devices, particularly as device encounters become more frequent thanks to the proliferation of new and unique products. Communication or even physical interaction with a device can expose a user to various threats, such as biometric theft or exploit of their own device. To address this, we propose a mechanism for verifying the integrity and trustworthiness of an IoT device before physical interaction or any significant communication has taken place.

\end{abstract}


%\listoftodos

\bstctlcite{IEEEexample:BSTcontrol}

\section{Introduction}

Internet of Things (IoT) devices, with specialised forms of human-computer interaction, pose challenges for establishing trust. For example, it is difficult to know whether a smart home assistant is listening to your conversation. Physically interacting with an unfamiliar device may pose risks, too, with devices harvesting biometric data without the victim knowing.

Many devices include trust mechanisms so that the manufacturer and interfacing devices can verify the operating integrity of the device. Users, meanwhile, must place implicit trust in these mechanisms. For example, Secure Boot asserts that the system trusts the Operating System that is being loaded, and by extension, so can the user.

In practice, rogue devices may be encountered, either via purpose-built IoT hardware, or through a breach of a presumed-trustworthy system. Users may have no easy way of establishing trust in a device prior to interacting with it.

In this paper, we propose a solution to this problem, extending trust from a user and their personal devices, towards nearby IoT devices, supported by a verification platform.
The system, dubbed ``Touch-Less Trust'' (TLT), allows users to gain insight into the nature and status of a device before interacting with it, regardless of its communication and physical interfaces or purpose.

This paper makes the following contributions:

\begin{itemize}
    \item Proposes a trust architecture that encompasses the user, IoT devices, manufacturers and a verification platform.
    \item Details the cryptographic methods for providing this trust efficiently, suitable for IoT communication technologies.
    \item Provides a threat model and explanation of how the TLT mechanism provides additional protections to these threats over conventional controls.
\end{itemize}

\Cref{sec:background} of this paper briefly explores the background literature that informs and motivates this work. \Cref{sec:architecture} details the proposed architecture for TLT, with implementation considerations discussed in \cref{sec:crypto}. A threat model and analysis are contributed in \Cref{sec:examples}. Finally, \cref{sec:conclusions} concludes this work and suggests future work in this area.

\section{Background}
\label{sec:background}

This work is motivated by the ``IoT Droplock''~\cite{KerrisonDroplocksiThings2022}, a malicious IoT device posing as a smart lock, which has been programmed to collect and send a victim's fingerprint data back to a malicious actor. The victim need only touch the device's fingerprint scanner to become compromised.

Technologies such as Secure Boot~\cite{wilkins2013uefi} address this to some extent, by utilising chains of trust against a root that is burned or otherwise fixed into a device, to verify the integrity of software and data that are being loaded by a system during startup. While Secure Boot is traditionally a technology seen in PCs, the principle also applies for embedded devices~\cite{8804799}, often under alternative names such as Trusted Boot and Verified Boot. This is similar in nature to the Public Key Infrastructure (PKI) implemented to secure Internet communications~\cite{rfc5280}, albeit typically controlled by manufacturers and not independent Certificate Authorities (CAs). These trust mechanisms focus mainly on trust between manufacturer and device, though, and not the user.

Another essential part of trust is the cryptography used, which poses challenges for resource-constrained IoT device. While some algorithms, such as those based around Elliptic Curve Cryptography (ECC) mitigate this somewhat~\cite{MallouliFatma2019ASoC}, a new challenge is presented by the prospect of quantum computers, with post-quantum algorithms further burdening devices~\cite{Fernandez-CaramesTiagoM.2020FPtP}.

Steiner et al. have surveyed attestation of device integrity in wireless sensor networks~\cite{10.1145/2988546}, which covers similar fundamental concerns to this paper.

\section{Architecture}
\label{sec:architecture}

In this section, the top-level architecture for TLT is explained, along with security properties that must be observed in order to ensure the efficacy of TLT. Generic cryptographic functions are used, with the notation described in \Cref{tab:notation}.

% Despite not messing with any margin sizes, EDAS things a page as a margin that's too small,
% so let's arm-wrestle with EDAS...
\begin{table*}
\vspace{-1.75\baselineskip}
\end{table*}
% /wrestling

\begin{table}
\caption{Notation for cryptographic operations and artefacts}
\begin{center}
\begin{tabular}{|E|L{0.575\linewidth}|}
\hline
\textbf{Notation} & \textbf{Meaning} \\
\hline
A, B, \ldots & Participant (Alice, Bob, \ldots). \\ \hline
A_p, A_s & \(A\)'s public and secret (private) keys. \\ \hline
%
%\mathbb{F}_\text{name}(\ldots) & Function, optional name and arguments. \\ \hline
\mathcal{M}_\textit{name} & Generic named message or data. \\ \hline
%\mathcal{P}_\textit{n}, \mathcal{C}_\textit{n},
\mathcal{H}_\textit{n}, \mathcal{S}_\textit{n}, \mathcal{D}_\textit{n} &
%\(\mathcal{P}\)laintext, \(\mathcal{C}\)iphertext,
\(\mathcal{H}\)ash, \(\mathcal{S}\)ignature or \(\mathcal{D}\)ocument message \(n\). \\ \hline
m' \leftarrow m_0 \concat %m_1 \concat%
    \cdots \concat m_n & Concatenation of messages. \\ \hline

r \leftarrow \mathbb{R}() & Random number generation. \\ \hline
k \leftarrow \mathbb{K}(\ldots) & Key agreement or derivation. \\ \hline
m' \leftarrow \mathbb{F}(X_y, m) & Function \(\mathbb{F}\), one of \(\mathbb{E}\)ncrypt, \(\mathbb{D}\)ecrypt, \(\mathbb{S}\)ign or \(\mathbb{V}\)erify on \(m\) using \(X\)'s key, \(y\). \\ \hline
% m' \leftarrow \mathbb{D}(B_x, m) & Decrypt \(m\) using \(B\)'s key \(x\). \\ \hline
m' \leftarrow \mathbb{H}(m) & Hash of message \(m\).
%with salt \(s\).
\\ \hline
% m' \leftarrow \mathbb{S}(A_x, m) & Sign \(m\) using \(A\)'s key, \(x\). \\ \hline
% \mathbb{V}(A_y, m) & Verify \(m\) using \(A\)'s key, \(y\). \\ \hline
\mathbb{P}(\mathcal{D}_\textit{l}:\mathcal{D}_\textit{i}:\cdots, \mathcal{D}_\textit{root}) & PKI-based verification of a chain of \(\mathcal{D}_x\) with a root of trust, \(\mathcal{D}_\textit{root}\).
\\ \hline
\end{tabular}
\label{tab:notation}
\end{center}
\end{table}

\subsection{Actors and components}

The TLT ecosystem comprises three types of actor:

\begin{itemize}
    \item the user, who wishes to establish trust in the devices they might interact with;
    \item the manufacturer, who makes devices that they want to people to trust; and
    \item the authority, which maintains central infrastructure to support and govern TLT.
\end{itemize}
There are a number of component devices and systems:
\begin{itemize}
    \item the user's own trusted device;
    \item an untrusted device that is being subject to scrutiny; and
    \item the TLT data store that maintains a record of devices and trust information.
\end{itemize}
Additional actors are envisaged, such as auditors, which may allow third party scrutiny and reporting. Such a need is motivated by wide-reaching supply-chain issues such as the PKfail incident affecting Secure Boot~\cite{pkfail}. However, these exceed the scope of this particular paper.

\subsection{Root of trust and the authority}

As with the PKI that underpins the security of TLS~\cite{rfc8446}, TLT relies on a root of trust that rests with an authority. This authority, \(T\), must have an asymmetric key pair and identifying information, \(\mathcal{M}_\textit{info}\), signed into a certificate, \(\mathcal{D}_\textit{root}\):
\begin{IEEEeqnarray}{rCl}
\IEEEyesnumber \label{eq:root}
\IEEEyessubnumber \label{eq:rootkey}
T_p, T_s & \leftarrow & \mathbb{K}_\text{asym}\left(\mathbb{R}()\right) \\
\IEEEyessubnumber \label{eq:rootdoc}
\mathcal{D}_\textit{tosign} & \leftarrow & \mathcal{M}_\textit{info} \concat T_p \\
\IEEEyessubnumber
\mathcal{D}_\textit{root} & \leftarrow & \mathcal{D}_\textit{tosign} \concat \mathbb{S}\left(T_s, \mathcal{D}_\textit{tosign}\right).
\end{IEEEeqnarray}

User-owned devices must trust this authority in order to use the TLT ecosystem. Conversely, manufacturers must seek the trust of the authority to be able to participate.

Best practices should be followed, such as using a Hardware Security Module (HSM) for key protection, and delegating the authority's signing operations to an intermediate certificate. For brevity, we assume this is considered in implementation, rather than detailed herein.

\subsection{Manufacturers}

A manufacturer must register with the TLT authority, after which they can register devices. By comparison with the web's PKI model, the TLT model considers manufacturers to be intermediate authorities, able to sign data only in relation to the devices that they produce, similar to DNS name constraints as defined in RFC5280~\cite[\S\,4.2.1.10]{rfc5280}.

A manufacturer, \(M\), generates keys \(M_p\) and \(M_s\), and the material needed for its own signing certificate, \(\mathcal{D}_\textit{mcrt}\), chained to the authority's certificate.
Assuming key and document composition is the same process as eq.~\ref{eq:root}, then the manufacturer's certificate is chained as:
\begin{IEEEeqnarray}{rCl}
\IEEEyesnumber \label{eq:mcrt}
\mathcal{D}_\textit{mcrt} & \leftarrow & \mathcal{D}_\textit{tosign} \concat \mathbb{S}\left(T_{s}, \mathcal{D}_\textit{tosign}\right) .
\end{IEEEeqnarray}

\subsection{Devices}

\subsubsection{Birth}
\label{sec:birth}

Upon the creation (or birth) of a device, a manufacturer must generate a UUID~\cite{rfc4122} for the device and burn this into the device. At the same time, the device must have an asymmetric key pair, either self-generated or injected and securely stored in the device. For device \(A\) this is:
\begin{IEEEeqnarray}{rCl}
\IEEEyesnumber \label{eq:devbirth}
\IEEEyessubnumber \label{eq:devuuid}
\mathcal{M}_\textit{uuid} & \leftarrow & \mathbb{R}_\textit{uuid}() \\
\IEEEyessubnumber \label{eq:devkey}
A_{p}, A_{s} & \leftarrow & \mathbb{K}_\text{asym}\left(\mathbb{R}()\right) .
\end{IEEEeqnarray}

Alongside the key and UUID may be other device-specific information such MAC address, but these may not be universally unique. TLT presumes a device is not initially identifiable or attributable to a particular manufacturer, and so a true UUID is required. The manufacturer can sign this information into a chained certificate \(\mathcal{D}_\textit{device}\):
\begin{IEEEeqnarray}{rCl}
\IEEEyesnumber \label{eq:device_birth}
\IEEEyessubnumber
\mathcal{D}_\textit{tosign} & \leftarrow & \mathcal{M}_\textit{dinf} \concat A_p \concat \mathcal{M}_\textit{uuid} \\
\IEEEyessubnumber
\mathcal{D}_\textit{dcrt} & \leftarrow & \mathcal{D}_\textit{tosign} \concat \mathbb{S}\left(M_s, \mathcal{D}_\textit{tosign}\right) .
\end{IEEEeqnarray}

Up to now, one could presume that all participants store and exchange the required keys, certificates and other materials needed to signing or verify documents. To reduce storage and communication burden on IoT devices, they need only store data essential to their operation. For example, in addition to a device's keys and UUID, it can store hashes of other documents or fragments therein, such as:
\begin{IEEEeqnarray}{rCl}
\IEEEyesnumber \label{eq:device_storage}
\mathcal{H}_\textit{dcrt} & \leftarrow & \mathbb{H}(\mathcal{D}_\textit{dcrt})
\end{IEEEeqnarray}
which should only be done if the device has seen the associated document verified its signature. For future exchanges, hashes can be extended from a digest such as \(\mathcal{H}_\textit{dcrt}\) with additional data and a new signature.

\subsubsection{Programming}

As with traditional firmware verification, the device must be able to verify the signature of firmware against the manufacturer's signing key. As the device is part of a TLT ecosystem, the signed firmware, \(\mathcal{D}_\textit{fw}\), must be presented to be verified against the TLT chain of trust:
% Not really an equation or assignment, though...
\begin{IEEEeqnarray}{l}
\IEEEyesnumber \label{eq:fwverif}
\mathbb{P}(\mathcal{D}_\textit{fw}:\mathcal{D}_\textit{mcrt}, \mathcal{D}_\textit{root}) .
\end{IEEEeqnarray}

Upon successful verification and activation, the device must sign a confirmation of installation by appending its UUID, any device specific installation information (e.g. storage slot used) and an additional signature to \(\mathcal{D}_\textit{fw}\):
\begin{IEEEeqnarray}{rCl}
\IEEEyesnumber \label{eq:proof_install}
\IEEEyessubnumber \label{eq:install_info}
\mathcal{D}_\textit{itsn} & \leftarrow & \mathcal{D}_\textit{fw} \concat \mathcal{M}_\textit{uuid} \concat \mathcal{M}_\textit{instinfo} \\
\IEEEyessubnumber
\mathcal{D}_\textit{inst} & \leftarrow & \mathcal{D}_\textit{itsn} \concat \mathbb{S}\left(A_s, \mathcal{D}_\textit{itsn} \right) .
\end{IEEEeqnarray}

% Despite not messing with any margin sizes, EDAS things a page as a margin that's too small,
% so let's arm-wrestle with EDAS...
\begin{table*}
\vspace{-1.75\baselineskip}
\end{table*}
% /wrestling

The manufacturer must be able to verify that the flash contents are correctly programmed~\cite{10.1145/2988546}. Future updates can be applied in a similar fashion.

\subsubsection{Configuration}

Beyond firmware updates, it may be beneficial for configuration or other state changes to also be signed into the device~\cite{10.1145/2988546}, similar to installation as in eq.~(\ref{eq:proof_install}).

\subsection{Users}

A user, with their own device, wishes to establish the trustworthiness of a new device. The device must have a means to broadcast its TLT UUID, for example over Bluetooth Low Energy (BLE).
With this UUID, the user's device can query the TLT data store to verify basic device information, such as model and purpose.

Following this, the user's device can issue a random challenge to the untrusted device, \( \mathcal{M}_\textit{ch} \leftarrow \mathbb{R}() \), which the device must include in a message containing identification, firmware and configuration integrity data, along with its own random data, signed by its private key:
\begin{IEEEeqnarray}{rCl}
\IEEEyesnumber \label{eq:user_challenge}
\IEEEyessubnumber
    \mathcal{H}_\textit{state} & \leftarrow & \mathbb{H}\left( \mathcal{D}_\textit{inst} \concat  \mathcal{D}_\textit{cfg} \concat \ldots \right) \\
\IEEEyessubnumber
    \mathcal{D}_\textit{rtosign} & \leftarrow & \mathcal{H}_\textit{state} \concat \mathcal{M}_\textit{ch} \concat \mathbb{R}() \\
\IEEEyessubnumber
    \mathcal{D}_\textit{resp} & \leftarrow & \mathcal{D}_\textit{rtosign} \concat \mathbb{S}\left( A_s, \mathcal{D}_\textit{rtosign} \right) .
\end{IEEEeqnarray}

The user's device can then query the TLT data store again for an entry keyed by \(\mathcal{H}_\textit{state}\), to verify the state of the device.

At this point, this information can be presented to the user for acceptance, or if use-case appropriate, automatically accepted by an application. Subsequently, the user's device might continue communication, or the user may now be confident to physically interact with it.

\section{Implementation considerations}
\label{sec:crypto}

This section explores implementation of TLT, including contemporary and future cryptography that may protect it.

\subsection{Communication} 

A mechanism must exist for communicating between the TLT-enabled device and the verifying device (i.e. IoT device and smartphone). BLE, for example, has advertising channel payloads of up to 31 bytes, and data payloads of up to 255 bytes. A 16-byte TLT UUID can fit in the former, while the latter can accommodate challenge/response messages. Bluetooth 5 allows for extended advertising channel payloads in BLE up to 1,650 bytes~\cite{woolley2021bluetooth}, accommodating larger signatures and other data, albeit with fragmentation.

\subsection{Current cryptography}

Currently, commonly used ECC keys are around 256 bits in length, which does not place a significant storage or communication burden on a device. Similarly, SHA-256 signatures are the same size, and are widely used. ECDSA and EdDSA signatures based on 256-bit keys are 512 bits in length. We would recommend these in preference to significantly larger RSA keys and signatures~\cite{MallouliFatma2019ASoC}.

By limiting device-side storage to keys, hashes, and operationally essential data, the overheads on the device are reduced. The TLT store centrally maintains the full-form of all artefacts, %such as signed firmwares, configuration states and intermediate trust chains,
so that they can be looked up when needed.
The verifying device is expected to have a higher bandwidth connection than the IoT device it is interacting with, and so is able to easily download such data where needed.

\subsection{Post-quantum considerations}

Post-Quantum Encryption (PQE) in IoT devices poses several challenges. Firstly, PQE algorithms may not be practical to run on embedded devices due to their CPU or memory constraints. Secondly, key, signature and other artefact lengths may be significantly longer than those used in conventional solutions.
Nevertheless, we have sought to define the TLT architecture to allow algorithms to be replaced in the future.

While NIST is working on standardisation of PQE algorithms (FIPS 203--205) and lightweight algorithms for IoT devices~\cite{turan2023status}, NIST recommends using strong conventional encryption on IoT devices~\cite{boutin2023nist}, leaving the majority of PQE concerns to larger systems that centralise larger amounts of data that are more critical in nature.

\section{Use cases and threat examples}
\label{sec:examples}

\begin{table*}
% EDAS: topmargins_a4: Upload failed: The top margin is 1.897 cm on page 4, which is below the required margin of 1.9 cm.
% Steve: 
\vspace{0.1\baselineskip}
% ...
\caption{Threats, assets and security controls in a TLT-enhanced IoT ecosystem}
\begin{center}
\begin{tabular}{|V|c|V|c|V|c|}
\hline
\multicolumn{2}{|c|}{\textbf{Threat Actors}} & \multicolumn{2}{|c|}{\textbf{Assets}} & \multicolumn{2}{|c|}{\textbf{Security Controls}}\\ \hline
\textbf{ID} & \textbf{Description} & \textbf{ID} & \textbf{Description}  & \textbf{ID} & \textbf{Description} \\ \hline
TA01 & Biometric harvesting & A01 & Biometrics & C01 & Proof of key possession \\ \hline
TA02 & Credential collection & A02 & PINs, passwords, other credentials & C02 & Proof of installed firmware \\ \hline
TA03 & Reverse exploit of app & A03 & Device key & C03 & Firmware update verification \\ \hline
TA04 & Reprogrammed device & A04 & Firmware & C04 & Secure / trusted boot \\ \hline
TA05 & Impostor device & A05 & Configuration & C05 & Proof of configuration \\ \hline
TA06 & Re-/mis-configured device & A06 & TLT chain & C06 & TLT check \\ \hline
\multicolumn{2}{|c|}{} & A07 & TLT artefacts & \multicolumn{2}{|c|}{} \\ \hline
\end{tabular}
\label{tab:threatmodel}
\end{center}
\end{table*}

\begin{figure}
  \centering
  %\fbox{
  \colorbox{white}{\includegraphics[width=0.96\linewidth,clip,trim=0cm 4.25cm 10.1cm 1cm]{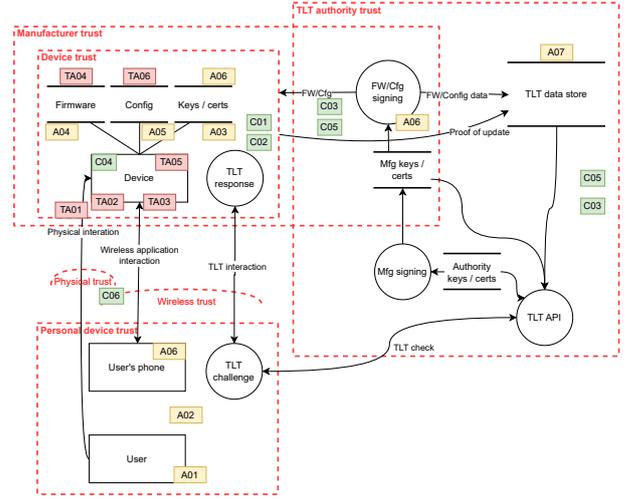}}
  %}
  \caption{Threat model for an IoT device with TLT controls added.}
  \label{fig:threatmodel}
\end{figure}

We have created a threat model, using M.~Henriksen's library for DrawIO\footnote{DrawIO library: \url{https://github.com/michenriksen/drawio-threatmodeling}}, presented in \cref{fig:threatmodel}. It focuses on aspects relevant to this work, and is not an exhaustive model.
Four main trust boundaries exist: the user and their personal device, the IoT device and components therein, the manufacturer that maintains the IoT device, the TLT authority.
Additionally, wireless and physical trust boundaries are crossed when the user or their device interacts with an IoT device.

The model annotates threat actors, \texttt{TA}, assets, \texttt{A}, and security controls, \texttt{C}, with corresponding identification numbers, described in \cref{tab:threatmodel}.
We now describe each threat actor, related assets, and how the controls mitigate those threats.

\begin{description}[font=\normalfont\itshape,leftmargin=1.0em]

    \item[\texttt{TA01} Biometric harvesting:]
        {A hacked or unknown device is physically interacted with and its biometric sensor collects a victim's data (\texttt{A01}). Control \texttt{C06} can be used to check integrity of firmware and configuration assets prior to physical interaction with the device.}
    
    \item[\texttt{TA02} Credential collection:]
        {Similar to \texttt{TA01} except the asset is \texttt{A02}, and \texttt{C06} is used prior to app communication.}
    
    \item[\texttt{TA03} Reverse exploit of app:]
        {This presumes a vulnerability in the user's phone app,
        \texttt{C06} can be used to prevent interaction with untrustworthy devices, which may
        avoid reaching a state where exploitation is possible.}
        
    \item[\texttt{TA04} Reprogrammed device:]
        {A combination of secure boot (\texttt{C04}) along with TLT checks (\texttt{C06}) and proof of installed firmware (\texttt{C02}) can detect this.}
        
    \item[\texttt{TA05} Impostor device:]
        {Lacking a trusted key (\texttt{A03}),
        the device will not be recognised along a TLT chain (\texttt{A06}) when performing a TLT check (\texttt{C06}).}
        
    \item[\texttt{TA06} Re-/mis-configured device:]
        {Similar to \texttt{TA04}, configuration updates are proven (\texttt{C05}) and checked the the user before proceeding (\texttt{C06}).}
\end{description}

\section{Conclusions and future work}
\label{sec:conclusions}

In this paper we have proposed an architecture through which trust in IoT devices can be established with minimal data exchange prior to full pairing, application communication or physical interaction. This addresses a security gap whereby the integrity and legitimacy of encountered IoT devices is difficult to establish.

The system, Touch-Less Trust (TLT), relies primarily on PKI and integration with existing trust mechanisms such as secure boot, along with firmware and other artefact storage, backed by an authority and participating manufacturers. Supported by the storage of artefacts in the TLT data store, seen and signed by devices and manufacturers, the TLT exchange with an untrusted device is limited to message sizes in the order of UUIDs, hashes and signatures, i.e. tens of bytes.

A proof-of-concept implementation of TLT, including a demonstration authority, manufacturer and devices, coupled with a user app and BLE-based implementation of the device checks, is a logical next step in this work. Beyond this, a performance evaluation and security audit can be performed, including a study of the impact on storage, communication and battery requirements. Handling encrypted firmwares, configurations and privacy-sensitive aspects should also be considered. Proving the installation of a firmware / configuration are also earmarked for future work.

\section*{Acknowledgements}

The author thanks the reviewers and committee of \mbox{TENCON 2024}, JCU's Belinda Lee and Ivy Lim for their administrative support and Dan Page for food for thought.

\bibliographystyle{IEEEtran}
\bibliography{touchless}

% Enfore empty line after bibliography
\ \\
{\noindent\footnotesize All URLs/DOIs were last followed on 27th September 2024.}

\end{document}